# DESIGN AND IMPLEMENTATION OF ENERGY EFFICIENT LIGHTWEIGHT ENCRYPTION (EELWE) ALGORITHM FOR MEDICAL APPLICATIONS

**Radhika Rani Chintala (Corresponding Author)**
*(Research Scholar, Department of Computer Science and Engineering, Koneru Lakshmaiah Education Foundation, Vaddeswaram, AP, India. Email: radhikarani_cse@kluniversity.in)*

**Narasinga Rao M R, Somu Venkateswarlu**
*(Research Supervisor, Department of Computer Science and Engineering, Koneru Lakshmaiah Education Foundation, Vaddeswaram, AP, India.)*

**ABSTRACT**

**Proportional to the growth in the usage of Human Sensor Networks (HSN), the volume of the data exchange between Sensor devices is increasing at a rapid pace. In HSNs, the sensors are either implanted or placed on the human body which are responsible for sensing the health data such as BP, Sugar level, Heart rate, etc. All this collected information is processed by the base station and is sent to the HSN servers via internet. Medical assistants who will have access to servers will analyze the health data of the patient and suggest appropriate medication. This will allow the doctors to have continuous monitoring on patient's health condition. Since the data is related to patient's health, it should be secured and confidential. In this paper, we have proposed an Energy Efficient Lightweight Encryption (EELWE) algorithm for providing the confidentiality of data at sensor level, particularly suitable for resource-constrained environments. Results obtained have proved that an EELWE consumes less energy relative to present lightweight ciphers and it supports multiple block sizes of 32-bit, 48-bit and 64-bit.**

**Keywords: Human Sensor Network, Security, Lightweight Encryption, Energy Consumption.**

## 1. INTRODUCTION

In HSNs, number of sensor nodes are involved, where communication between the nodes is done within a little distance using wireless links/connections. Conventional encryption algorithms don't fit for these types of environments, because they demand more capacity resources. For all these connected devices, information security is evidently necessary [1]. As the available resources in sensor devices are limited and due to the huge demand of security in resource-constrained environments, the need for lightweight encryption algorithms became essential [2].

Asymmetric and Symmetric are two types of cryptographic security algorithms [3]. Asymmetric type algorithms are the choice for non-repudiation and authentication. In these algorithms, encoding and decoding of data is done using separate keys. Symmetric type algorithms are mostly used for privacy and confidentiality. In these algorithms, same key is utilized for both encoding and decoding. Statistical functions are used in each round for purpose of including confusion and diffusion.

Lightweight cryptography is block cipher-based encoding method [4], used in resource constrained devices. It provides confidentiality in high-speed and lightweight environments such as sensor devices.

Criteria for Lightweight block ciphers include [5,6]:
- Tiny block size of <= 64-bit
- Tiny key size of <= 80-bit
- Round function with simple logic

  ➢ Less complexity in key scheduling.

Though better security is provided by many existing algorithms but are consuming relatively more battery [7]. Thus, causing the tiny sensor devices to exhaust fast. Hence, there is a need for a new algorithm that ensures better or same level of security as the existing algorithms, by consuming less energy.

The organization of the remaining part of the paper is given here: Review of different lightweight encryption algorithms is briefed in section 2. The full narration of the proposed EELWE algorithm is available in section 3. Sections 4 discusses about parameters needed for calculating energy consumption of EELWE. Metric for Security vs Energy consumption is calculated in Section 5. Finally, the results obtained after simulation of EELWE using Xilinx and results of energy consumption are shown in section 6.

## 2. LITERATURE REVIEW

Regardless of the significant amount of work done in this field of security in HSN, only a few literatures came up with suitable framework for calculating the energy requirements of the existing algorithms.

AES is a popular block cipher developed by Vincent Rijmen and Joan Daemen [8], which is a subcategory of Rijndael block cipher. It follows an SPN structure. It works with three key lengths of size 128-bit, 192-bit, 256-bit and undergoes the iterations of 10,12,14 rounds, respectively.

DESL and DESXL [9] are the two lightweight forms of the standard DES. They follow Feistel structure. DESL which is same as DES, takes a 64-bit block plaintext, a key of size 56-bit and undergoes the iterations of 16 rounds. All the rounds make use of a single S-box. The linear property of S-box is the drawback of DES, which is been reinforced in DESL by utilizing a single non-linear S-box eight times. DESXL takes a key size of 184-bit and it uses a key whitening technique to strengthen the security.

HIGHT is a hardware-oriented lightweight encryption protocol [10]. The operations of HIGHT can be performed using a small 8-bit processor and it follows Feistel structure. It takes a 64-bit block data, a key of size 128-bit and undergoes the iterations of 32 rounds. The round function consists of simple logical and arithmetic operations such as bitwise left rotation, XOR, addition and subtraction with mod2⁸.

IDEA is a traditional symmetric block cipher [11] that is developed by Xuejia Lai and James Massey in 1991. It was proposed as a successor of DES. It takes a 64-bit block data, a key of size 128-bit and undergoes the iterations of 8 similar rounds, which is followed by a half output round. Thus, it undergoes a total of 8.5 rounds.

KLEIN is a lightweight block cipher [12] that was proposed by Gong et al. in 2011 during RFIDSec. It follows SPN structure and has an innovative structure that uses the combination of 4-bit S-boxes and Mix column transformations of AES. It takes a 64-bit block data, a key of variable lengths 64-bit, 80-bit, 96-bit and undergoes the iterations of 12, 16, 20 rounds, respectively.

LBLOCK is a lightweight encryption protocol [13] that follows the Feistel structure. It takes a 64-bit block data, a key of size 80-bit and undergoes the iterations of 32 rounds. Efficient implementation of LBLOCK is possible in both hardware and software.

LED is a lightweight block cipher [14] that was designed keeping hardware compaction in mind. Its main advantage is that the key can be provided by a user rather than using a fixed key. It follows SPN structure. It is also similar to AES and makes use of S-boxes, mix columns and shift rows. It takes a 64-bit block data, a key of size 64-bit, 128-bit and undergoes the iterations of 32, 48 rounds respectively and also re-utilize the S-box of the block cipher PRESENT.

mCrypton falls under the lighter version of Crypton block cipher [15] that is specially designed for devices having limited resources. It follows SPN structure. It takes a 64-bit block data and works with three key lengths of size 128-bit, 96-bit, 64-bit and undergoes the iterations of 32 rounds. Key scheduling is done in two steps: 1) Using S-box for generating round key, 2) Using word wise and bitwise rotations for updating key variable.

PICCOLO is a lightweight encryption method that follows Feistel structure [16]. It takes a 64-bit block data, and works with two different key lengths of size 80-bit, 128-bit and undergoes the iterations of 25, 31 rounds respectively. The round logic of Piccolo takes three XOR, one XNOR and four NOT gates which is equal to 12 GE and this is the earliest block cipher whose hardware implementation is done with not more than 1000 gates.

PRESENT falls under the ultra-lightweight block cipher [17] that follows the SPN structure. It takes a 64-bit block data and works with two different key lengths of size 80-bit, 128-bit and undergoes the iterations of 31 rounds. The round function of PRESENT contains 3 steps: 1) XOR logic for AddRoundKey, 2) 4-bit S-box is used for 16 times parallelly in S-box layer, 3) Bitwise permutations in P-Layer.

TEA is Tiny Encryption Algorithm [18] developed by D. Wheeler and N. Needham. XTEA is an extension for TEA with 64-bit block size and uses 128-bit key [19]. It follows Feistel structure and uses simple F-function. Since the round function is simple, it undergoes 64 no. of rounds to increase the security.

SEA is relatively cheap encryption protocol [20] designed explicitly for hardware with restricted no of instructions. Likewise, this is an adaptable protocol w.r.t. Size of text, length of key & word size of the processor. It follows Feistel structure, a key of variable lengths and undergoes varying number of rounds.

The comparison of the existing lightweight block ciphers is presented in Table 1.

Table 1. Lightweight block ciphers comparison

| Name of Algorithm | Key Size | Block Size | No of Rounds | Structure |
|---|---|---|---|---|
| AES | 128,192,256 | 128 | 10,12,14 | SPN |
| DESXL | 184 | 64 | 16 | Feistel |
| HIGHT | 128 | 64 | 32 | Feistel |
| IDEA | 128 | 64 | 8 | Lai-Massey |
| KLEIN | 64, 80, 96 | 64 | 12,16,20 | SPN |
| LBLOCK | 80 | 64 | 32 | Feistel |
| LED | 64, 128 | 64 | 32, 48 | SPN |
| mCrypton | 64, 96, 128 | 64 | 32 | SPN |
| Piccolo | 80, 128 | 64 | 25, 31 | Feistel |
| PRESENT | 80, 128 | 64 | 31 | SPN |
| XTEA | 128 | 64 | 64 | Feistel |
| SEA | "on-the-fly" | 128 | variable | Feistel |

### 3. METHODOLOGY

**3.1 Structure Used:**
The algorithm that is proposed follows Feistel structure [5]. Feistel structure systems offer both encryption and decryption with less expense [21]. The main advantage of using Feistel structure is that the encoding and decoding process use the same form of structure.

**3.2 Number of Rounds:**
More rounds mean greater security [6]. For a protected block cipher, there ought to be no attack that is faster than an exhaustive key searching like brute force. As a thorough key search takes much longer for a bigger key size, a hypothetical attacker must manage the cost of more work to "break" the cipher. Therefore, one must increment the number of rounds for increasing security of a cipher.

### 3.3 Lightweight Algorithm Functionality:

The organization of the lightweight encryption algorithm is presented in Fig.1. It contains two main blocks namely Key scheduling block and round function block with T rounds. Each round takes n-bit data that is generated by the previous round, performs the encoding using the sub-key, and generates an n-bit ciphertext, which in turn is given as an input to the next round. The key schedule function will take the master key as input & expands it into sub-keys, where each sub-key is given to one round.

The implementation of the lightweight block cipher algorithm shown in Fig.2 contains the blocks, namely registers, Overhead logic, and the round's function.

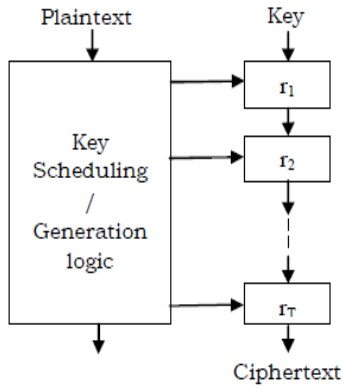

Fig.1 Encoding Algorithm

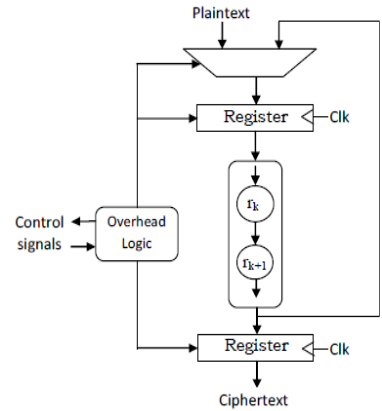

Fig.2 Implementation of Encryption Algorithm

Registers are used to save the initial data, intermediate data, and the final data. Overhead logic is used to generate the sub-keys. Round's function is used to implement ri rounds. Lightweight block ciphers usually have more significant no. of rounds with simple operations & simple key schedule functionality [22].

### 3.4 Proposed EELWE Algorithm

EELWE algorithm is shown presented in Fig.3.

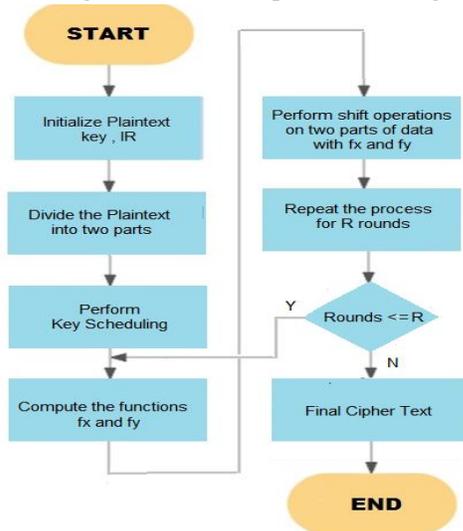

The proposed EELWE block cipher is in three variants: EELWE32, EELWE48, and EELWE64 with blocks of sizes 32-bit, 48-bit, and 64-bit. All these ciphers have a fixed key of length 80-bit and go through 254 rounds. They perform key scheduling to provide keys for 254 rounds and make all the variants use the same nonlinear round functions.

The flowchart representing the functionality of the

Fig.3 Flowchart of EELWE

The notations used in the algorithm are given in Table 2.

Table 2. Notations used to explain EELWE Algorithm.

| Notation | Description |
|---|---|
| P1, P2 | First part and Second part of Plaintext |
| Pa, Pb | First part and Second part of P1 |
| Pc, Pd | First and Second part of P2 |
| ai, bi, ci, di | Bit positions of Pa, Pb, Pc, Pd respectively |
| fx | Round function on $1^{st}$ part of data using ai and bi values, IR and subkey kx |
| fy | Round function on $2^{nd}$ part of data using ci and di values and subkey ky |
| $\oplus$, & | Logical XOR and AND operation |
| IR | Irregular Update rule |
| ki | $i^{th}$ bit of the key |
| kx | Subkey applied on P1 where kx = $k_{2i}$ |
| ky | Subkey applied on P2 where ky = $k_{2i+1}$ |

EELWE32 is the basic algorithm of the EELWE family with a plaintext of size 32-bit and produces a ciphertext of the size of 32-bit. The given plaintext is split up into two parts of 13-bit & 19-bit. Register P1 is loaded with 13-bit data and register P2 is loaded with 19-bit data. The $12^{th}$ bit of P1 represents the MSB of the plaintext, and the $0^{th}$ bit of P2 represents the LSB of plaintext.

**Round Function:**
P1 data is further split up into 2 parts Pa and Pb, each of 6-bits and 7-bits, respectively. The round function is performed on Pa and Pb using the following nonlinear functions.

$$Ta = (Pa[a1] \oplus Pa[a2]) \oplus kx$$
$$Tb = (Pb[b1] \& Pb[b2]) \oplus (Pb[b3] \& IR[i])$$
$$fx = Ta \oplus Tb$$

Ta is computed on Pa data using subkey kx and ai values, and Tb is computed on Pb data using IR (irregular) update rule and bi values. Similarly, Tc and Td are computed based on the below rules.

$$Tc = (Pc[c1] \oplus Pc[c2]) \oplus ky$$
$$Td = (Pd[d1] \& Pd[d2]) \oplus (Pd[d3] \& Pd[d4])$$
$$fy = Tc \oplus Td$$

ai, bi, ci, and di represent the random bit positions of Pa, Pb, Pc, and Pd, respectively. Now left shift operation is performed on P1 and P2 by inserting fy and fx into LSBs of P1 and P2, respectively.

$$P1 = shl(P1, fy)$$
$$P2 = shl(P2, fx)$$

P1 and P2 are updated with new data in each round. After the completion of 254 rounds, P1 and P2 contents denote the final ciphertext where the $12^{th}$ bit of P1 represents the MSB of the ciphertext, and the $0^{th}$ bit of P2 represents the LSB of the ciphertext.

**Key Scheduling:**
An 80-bit key is expanded using the key scheduling, where two subkeys (kx and ky) are used in each round. kx = $k_{2i}$ and ky = $k_{2i+1}$, where i denotes the $i^{th}$ round. The subkey kj (which may denote $k_{2i}$ or $k_{2i+1}$) is given as:

$$kj = kj \quad \text{for } j = 0 \text{ to } 79$$
$$kj = k_{j-80} \oplus k_{j-61} \oplus k_{j-50} \oplus k_{j-13}$$
$$\text{for } j = 80 \text{ to } 253$$

In EELWE32/48/64, the plaintext is divided into 13 and 19 / 19 and 29 / 25 and 39-bit blocks. P1 of EELWE32/48/64 is further divided into 6 and 7 / 8 and 11 / 10 and 17 bit blocks with a = {5,2} / {7,3} / {9,5} and b = {6,3,1} / {10,6,4} / {16,11,7}. Similarly, P2 of EELWE32/48/64 is further divided into 8 and 11 / 12 and 17 / 14 and 23 bit blocks with c = {7,3} / {11,5} / {13,7} and d = {10,7,5,1} / {16,12,8,3} / {22,17,11,5}. For each iteration, the round function is performed twice in EELWE48 and thrice in EELWE64.

**4. ENERGY CONSUMPTION OF PROPOSED ALGORITHM:**

To calculate the energy consumed by an encryption algorithm, the three primary parameters required are Time, Area, and Power [23].

The following equation gives the time required for encrypting one block data:
$$T_B = ((r_T/r_i) + C_0) \times (D_R + r_i \times (t_0 + t_n \times n))$$

(1)

Area required for encrypting one block data is given by the following equation:
$$A_D = r_i^{(gn \times n + g0)} \times A_1 + g_b \times n + A_0$$

(2)

Power required for encrypting one block data is given by the following equation:
$$P = ((P_d \times r_i + P_i) \times A_D) / CT$$

(3)

Based on the above parameters, the energy consumed for encrypting one block data is given by the following equation:
$$E_B = T_B \times P$$

(4)

To have a fair comparison among the block ciphers, it is better to consider the energy per bit rather than energy per block. Energy per bit is calculated using the following equation:
$$E_1 = E_B / n$$

(5)

Different parameter values for energy consumption obtained for EELWE32, EELWE48, and EELWE64 are presented in Fig.4, Fig.5, and Fig.6, respectively.

| 32-bit Block (Plaintext = 5742414E) | | No of Rounds | | | | | | | | |
|---|---|---|---|---|---|---|---|---|---|---|
| S. No | No of Rounds (r)----> | 1 | 2 | 4 | 8 | 16 | 32 | 64 | 128 | 254 |
| 1 | Cipher Text in HEX | AE8C829D | 5D11053A | 745414E9 | 45414E98 | 4106982E | 6DC6913E | 349BCC76 | 35490E7A | 53E0848A |
| 2 | CB Cycles per Block | 256 | 129 | 66 | 34 | 18 | 10 | 6 | 4 | 3 |
| 3 | Tr1 Time for One Round | 0.0001459 | 0.0001459 | 0.0001459 | 0.0001459 | 0.0001459 | 0.0001459 | 0.0001459 | 0.0001459 | 0.0001459 |
| 4 | CT Time for One Cycle | 0.0141459 | 0.0142917 | 0.0145834 | 0.0151669 | 0.0163338 | 0.0186675 | 0.023335 | 0.0326701 | 0.0510484 |
| 5 | TB Time for One Block (ns) | 3.6213402 | 1.8436319 | 0.962507 | 0.5156739 | 0.2940077 | 0.1866752 | 0.1400102 | 0.1306803 | 0.1531453 |
| 6 | Throughput | 8.8365076 | 17.357044 | 33.24651 | 62.054719 | 108.84069 | 171.42074 | 228.55471 | 244.87237 | 208.95186 |
| 7 | Ar Area of Implemented R rounds | 30 | 53.006251 | 93.655421 | 165.47742 | 292.37793 | 516.59526 | 912.75926 | 1612.7315 | 2831.201 |
| 8 | AD Design Area(GE) | 206 | 229.00625 | 269.65542 | 341.47742 | 468.37793 | 692.59526 | 1088.7593 | 1788.7315 | 3007.201 |
| 9 | Power P (pW) | 14067.437 | 16536.46 | 21522.968 | 32150.961 | 56089.181 | 111750.08 | 239074.05 | 511815.78 | 1040564.6 |
| 10 | Eblock Energy per Block | 50942.976 | 30487.144 | 20716.008 | 16579.412 | 16490.65 | 20860.969 | 33472.815 | 66884.25 | 159357.6 |
| 11 | Eb Energy per Bit (pJ/bit) | 1591.968 | 952.72325 | 647.37525 | 518.10662 | 515.33282 | 651.90529 | 1046.0255 | 2090.1328 | 4979.9249 |

Fig.4 Parameter values obtained for EELWE32

| 48-bit Block (Plaintext = 5742414e3438) | | No of Rounds | | | | | | | | |
|---|---|---|---|---|---|---|---|---|---|---|
| S. No | No of Rounds (r)----> | 1 | 2 | 4 | 8 | 16 | 32 | 64 | 128 | 254 |
| 1 | Cipher Text in HEX | 5d0965 38d0e2 | 742594 e34389 | 42592e 343891 | 593db4 3891fb | a158d1 fb2f4b | 3f7846 2baa44 | d34c48 bd8631 | 176e9a 3837ba | 36b0a9 4934a6 |
| 2 | CB Cycles per Block | 256 | 129 | 66 | 34 | 18 | 10 | 6 | 4 | 3 |
| 3 | Tr1 Time for One Round | 0.0001473 | 0.0001473 | 0.0001473 | 0.0001473 | 0.0001473 | 0.0001473 | 0.0001473 | 0.0001473 | 0.0001473 |
| 4 | CT Time for One Cycle | 0.0141473 | 0.0142946 | 0.0145892 | 0.0151783 | 0.0163566 | 0.0187133 | 0.0234266 | 0.0328531 | 0.0514117 |
| 5 | TB Time for One Block | 3.6217062 | 1.8440008 | 0.9628846 | 0.5160629 | 0.2944195 | 0.1871328 | 0.1405594 | 0.1314125 | 0.154235 |
| 6 | Throughput | 13.253422 | 26.030357 | 49.850213 | 93.011921 | 163.03267 | 256.50233 | 341.49273 | 365.26211 | 311.21345 |
| 7 | Ar Area of Implemented R rounds | 30 | 55.472106 | 102.57182 | 189.66249 | 350.69926 | 648.46756 | 1199.062 | 2217.1499 | 4071.2502 |
| 8 | AD Design Area | 254 | 279.47211 | 326.57182 | 413.66249 | 574.69926 | 872.46756 | 1423.062 | 2441.1499 | 4295.2502 |
| 9 | Power P | 17343.534 | 20176.543 | 26055.619 | 38918.012 | 68725.103 | 140428.2 | 311260.8 | 694602.8 | 1475760.6 |
| 10 | Eblock Energy per Block | 62813.184 | 37205.563 | 25088.553 | 21084.141 | 20234.012 | 26278.723 | 43750.619 | 91279.477 | 227613.9 |
| 11 | Eb Energy per Bit(pJ/bit) | 1308.608 | 775.11589 | 522.6782 | 439.25295 | 421.54191 | 547.47339 | 911.47124 | 1901.6558 | 4741.9562 |

Fig.5 Parameter values obtained for EELWE48

| S. No | 64-bit Block (Plaintext = 5742414E4B4C4345)<br>No of Rounds (r)----> | 1 | 2 | 4 | 8 | 16 | 32 | 64 | 128 | 254 |
|---|---|---|---|---|---|---|---|---|---|---|
| 1 | Cipher Text in HEX | ba120b725 a621a2c | d0905b92d 310d164 | 2416fc34c4 345936 | 6fc3534345 9364b1 | 2b884e64b 17d41f7 | a1a57160c0 6e9bd2 | 2b6e577f7c c1fa3f | 6be00fcf06 8f7746 | 63876ffab1 a27290 |
| 2 | CB Cycles per Block | 256 | 129 | 66 | 34 | 18 | 10 | 6 | 4 | 3 |
| 3 | Tr1 Time for One Round | 0.0001487 | 0.0001487 | 0.0001487 | 0.0001487 | 0.0001487 | 0.0001487 | 0.0001487 | 0.0001487 | 0.0001487 |
| 4 | CT Time for One Cycle | 0.0141487 | 0.0142974 | 0.0145949 | 0.0151898 | 0.0163795 | 0.018759 | 0.0235181 | 0.0330362 | 0.0517749 |
| 5 | TB Time for One Block | 3.6220723 | 1.8443698 | 0.9632621 | 0.5164518 | 0.2948314 | 0.1875904 | 0.1411085 | 0.1321446 | 0.1553246 |
| 6 | Throughput | 17.669443 | 34.7002 | 66.4409 | 123.92249 | 217.07324 | 341.16884 | 453.55176 | 484.31779 | 412.04023 |
| 7 | Ar Area of Implemented R rounds | 30 | 58.052673 | 112.3371 | 217.38229 | 420.6541 | 814.00317 | 1575.1687 | 3048.0917 | 5854.4336 |
| 8 | AD Design Area | 302 | 330.05267 | 384.3371 | 489.38229 | 692.6541 | 1086.0032 | 1847.1687 | 3320.0917 | 6126.4336 |
| 9 | Power P | 20618.968 | 23823.451 | 30652.419 | 46007.173 | 82714.965 | 174371.48 | 402451.74 | 939462.01 | 2090151.1 |
| 10 | Eblock Energy per Block | 74683.392 | 43939.252 | 29526.313 | 26760.489 | 24386.966 | 32710.415 | 56789.353 | 124144.87 | 324651.97 |
| 11 | Eb Energy per Bit(pJ/bit) | 1166.928 | 686.55082 | 461.34864 | 418.13264 | 381.04634 | 511.10024 | 887.33364 | 1939.7636 | 5072.687 |

Fig.6 Parameter values obtained for EELWE64

## 5. SECURITY Vs. ENERGY CONSUMPTION

Key length plays an important role in defining the security level of an encryption algorithm [24]. If a processor of P Flops encrypts a block of data in C clock cycles, then the number of encryption operations(O) that can be carried out in M years using N systems can be calculated as:

$$O = (P / C) * 60 * 60 * 24 * 365 *$$

(6)

From eq.6, we can obtain the approximate key length that is to be maintained by an encryption algorithm and based on this an 80-bit key is used by an EELWE algorithm. According to the Moore's Law [25], the year Y(γ) till which the cipher will provide adequate protection is calculated as:

$$Y(\gamma) = \text{Proposed Year} + 3(\gamma - 56)$$

(7)

From eq.7, it is proved that EELWE is secured till the 2317 with security level(γ) of 254.

To measure the security vs energy consumption, a metric called MSEC is used which is calculated as:

$$MSEC = (\text{No. of Secured years left}) / (\text{Normalized energy})$$

(8)

It is shown that an EELWE has exhibited better MSEC value than the existing algorithm.

## 6. RESULTS

The proposed EELWE algorithm is implemented in Xilinx. Xilinx is an s/w tool that enables you to synthesize and simulate the HDL (Hardware Description Language) design [26]. The encryption and decryption results of EELWE32, EELWE48 and EELWE64 are presented in Fig.7, Fig.8 and Fig.9, respectively

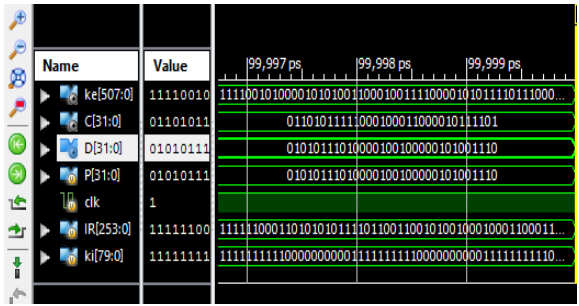
Fig.7 Implementation results for EELWE32

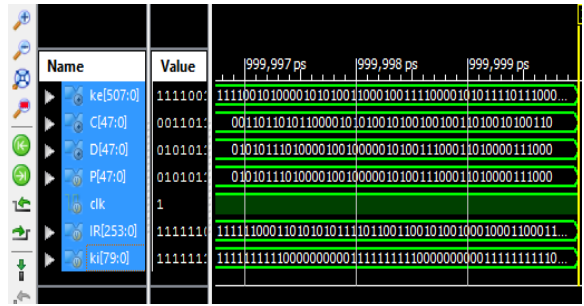
Fig.8 Implementation results for EELWE48

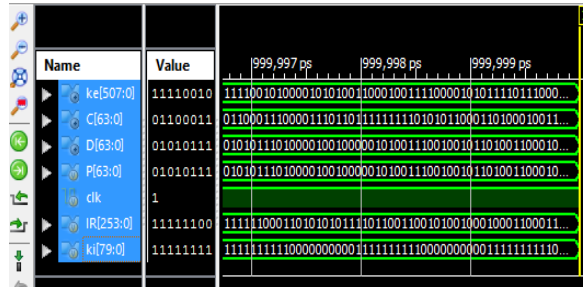
Fig.9 Implementation results for EELWE64

The energy consumption of different existing lightweight encryption algorithms and the proposed algorithm is shown in the below Table 3. When compared to the existing algorithms, EELWE is consuming less energy.

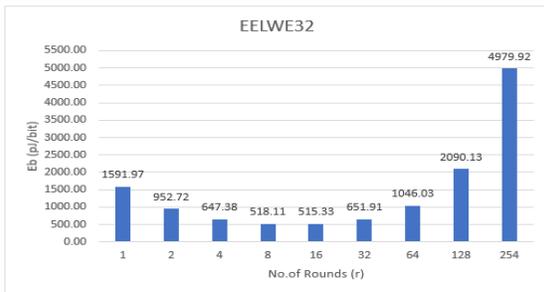
Fig.10 Energy consumption for EELWE32

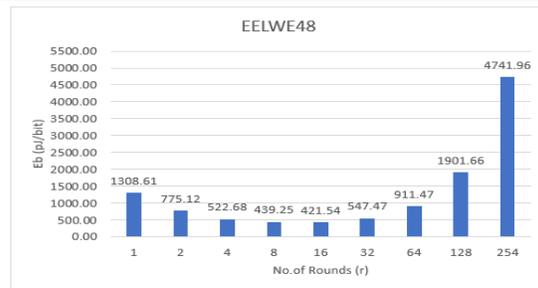
Fig.11 Energy consumption for EELWE48

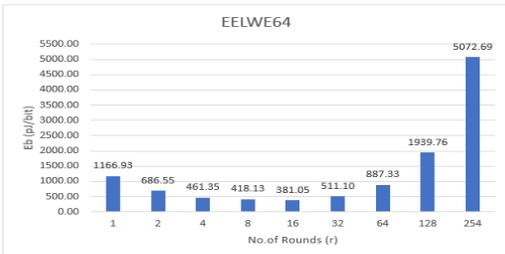
Fig.12 Energy consumption for EELWE64

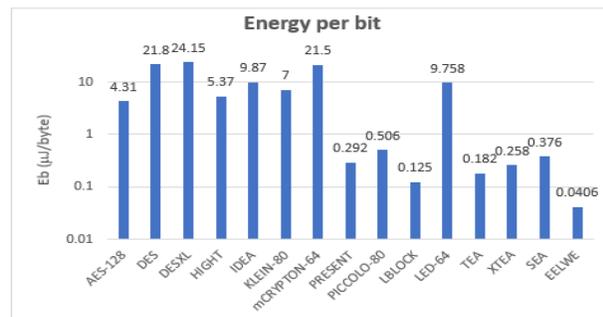
Fig.13 Energy consumption of various LEW algorithms

The graphs show that the minimum energy is consumed at 16 rounds in all the three versions. The comparison of energy consumed by various algorithms w.r.t to their block sizes is shown in Table 3 and the graph representation is shown in Fig.13.

Table 3. Comparison of Energy consumed by various algorithms.

| Algorithm | Block Size(bits) | Energy consumption for Encryption (µJ/byte) |
|---|---|---|
| AES-128 | 128 | 4.31 |
| DES | 64 | 21.8 |
| DESXL | 64 | 24.15 |
| HIGHT | 64 | 5.37 |
| IDEA | 64 | 9.87 |
| KLEIN-80 | 64 | 7.0 |
| mCrypton-64 | 64 | 21.5 |
| PRESENT | 64 | 0.292 |
| Piccolo-80 | 64 | 0.506 |
| LBLOCK | 64 | 0.125 |
| LED-64 | 64 | 9.758 |
| TEA | 64 | 0.182 |
| XTEA | 64 | 0.258 |
| SEA | 96 | 0.376 |
| EELWE | 64 | 0.0406 |

MSEC value of EELWE and different existing algorithms are shown in Fig.14.

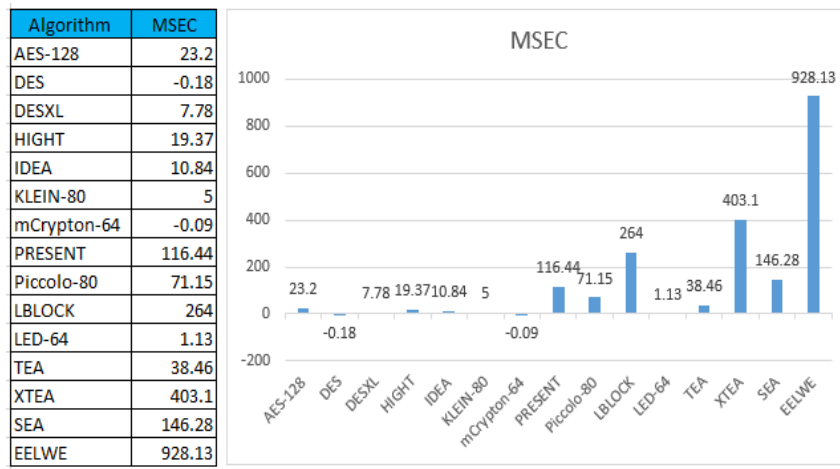

Fig.14 MSEC value of different algorithms

## 7. CONCLUSION

A new Lightweight encryption algorithm EELWE is presented in this paper. When simulated in the Xilinx environment, EELWE has consumed comparably less energy than several existing lightweight ciphers. It has supported three different block sizes that are suitable for resource constraint environments such as HSNs. From the observations, it is concluded that EELWE64 obtains the optimized energy consumption with an 80-bit key and for 16 rounds i.e., 0.00305 µJ/byte. But, to have better security, it is suggested to undergo 254 rounds which consumes an energy of 0.0406 µJ/byte. Even this value is far less than energy consumed by other algorithms. Also, EELWE has exhibited better MSEC value compared to existing algorithms.

## 8. REFERENCES

[1]. Radhika Rani Chintala, Nrasinga Rao M R, Somu Venkateswarlu, "A Review on Security Issues in Human Sensor Networks for Healthcare Applications", International Journal of Engineering & Technology, Vol. 7, No. 2.32, pp. 269-274, 2018.

[2]. Ch.Radhika Rani, Lakku Sai Jagan, Ch. Lakshmi Harika, V.V. Durga Ravali, "Lightweight Encryption Algorithms for Wireless Body Area Networks", International Journal of Engineering & Technology, Vol. 7, No. 2.20, pp. 64-66, 2018.

[3]. Michelle S. Henriques, Nagaraj K. Vernekar, "Using symmetric and asymmetric cryptography to secure communication between devices in IoT," International Conference on IoT and Application (ICIOT), IEEE, 2017.

[4]. Nikita Arora1 and Yogita Gigras, "Block and Stream Cipher Based Cryptographic Algorithms: A Survey, International Journal of Information and Computation Technology, ISSN 0974-2239 Volume 4, Number 2, pp. 189-196, 2014.

[5]. Deepti Sehrawat and Nasib Singh Gill, "Lightweight Block Ciphers for IoT based applications: A Review," International Journal of Applied Engineering Research, ISSN 0973-4562, Volume 13, Number 5, pp. 2258-2270, 2018.

[6]. George Hatzivasilis, Konstantinos Fysarakis, Ioannis Papaefstathiou, and Charalampos Manifavas, "A review of lightweight block ciphers," Journal of Cryptographic Engineering, volume 8, pp. 141–184, 2018.

[7]. J. Toldinas, R. Damasevicius, A. Venckauskas, T. Blazauskas, J. Ceponis, "Energy Consumption of Cryptographic Algorithms in Mobile Devices," in Elektronika IR Elektrotechnika, ISSN 1392–1215, Vol. 20, No. 5, 2014.

[8]. Joan Daemen, Vincent Rijmen, "AES Proposal: Rijndael", The Rijndael Block Cipher, Dec 2012.

[9]. Soufiane Oukili, Seddik Bri, "High Throughput FPGA Implementation of Data Encryption Standard with Time Variable Sub-Keys", International Journal of Electrical and Computer Engineering, Vol. 6, No. 1, 2016.

[10]. Bassam Jamil Mohd, Thaier Hayajneh, Zaid Abu Khalaf and Khalil Mustafa Ahmad Yousef, "Modeling and optimization of the lightweight HIGHTblock cipher design with FPGA implementation", Security and Communication Networks, Vol. 9, No. 13, pp. 2200-2216, 2016.

[11]. Pushpalatha G S , Harshitha N G , Rashmi C , Rashmi P K , Preksha S, "Performance Analysis of Idea Algorithm on FPGA for Data Security", International Journal for Research in Applied Science & Engineering Technology (IJRASET), Vol. 7, No. 5, pp. 2635 – 2638, 2019.

[12]. Pulkit Singh, B. Acharya, R. K. Chaurasiya, "High Throughput Architecture for KLEIN Block Cipher in FPGA", Annual Information Technology, Electromechanical Engineering and Microelectronics Conference (IEMECON), 2019.

[13]. Aljazeera.K.R, Nandakumar.R, Ershad.S.B, "Design And Characterization of LBlock Cryptocore", International Conference on Signal Processing, Communication, Power and Embedded System (SCOPES), 2017.

[14]. Mohammed Al-Shatari, Fawnizu Azmadi Hussin, Azrina Abd Aziz, Gunawan Witjaksono , Mohd Saufy Rohmad , Xuan-Tu Tran, "An Efficient Implementation of LED Block Cipher on FPGA", International Conference of Intelligent Computing and Engineering (ICOICE), 2019.

[15]. Chae Hoon Lim, Tymur Korkishko, "mCrypton – A Lightweight Block Cipher for Security of Low-Cost RFID Tags and Sensors", in Information Security Applications, vol. 3786 of LNCS, pp. 243–258, Springer Berlin Heidelberg, 2006.

[16]. Ayoub Mhaouch, Wajdi Elhamzi, Mohamed Atri, "Lightweight Hardware Architectures for the Piccolo Block Cipher in FPGA", International Conference on Advanced

Technologies for Signal and Image Processing (ATSIP), 2020.

[17]. Hengameh Delfan Azari, Dr. Prashant V Joshi, "An Efficient Implementation of Present Cipher Model with 80 bit and 128 bit key over FPGA based Hardware Architecture", International Journal of Pure and Applied Mathematics, Vol. 119, No. 14, pp. 1825-1832, 2018.

[18]. Kiran Kumar.V.G, Sudesh Jeevan Mascarenhas, Sanath Kumar, Viven Rakesh J Pais, "Design And Implementation Of Tiny Encryption Algorithm", International Journal of Engineering Research and Applications, Vol. 5, No. 6, pp. 94 – 97, 2015.

[19]. Jiazhe Chen, Meiqin Wang, Bart Preneel, "Impossible Differential Cryptanalysis of the Lightweight Block Ciphers TEA, XTEA, and HIGHT," International Conference on Cryptology in Africa, AFRICACRYPT, pp. 117-137, 2012.

[20]. K.Ashok Kumar, G.Senthil Kumar, S. Poonguzhali, "FPGA Implementation of Scalable Encryption Algorithm Using Veriloghdl With Xilinx Spartan-3", Indian Journal of Applied Research, Vol. 5, No. 2, 2015.

[21]. Taizo SHIRAI Kiyomichi ARAKI, "On Generalized Feistel Structures Using the Diffusion Switching Mechanism," IEICE TRANSACTIONS on Fundamentals of Electronics, Communications, and Computer Sciences, Vol.E91-A No.8 pp.2120-2129, 2008.

[22]. B. J. Mohd, T. Hayajneh, and A. V. Vasilakos, "A survey on lightweight block ciphers for low-resource devices: Comparative study and open issues," Journal of Network and Computer Applications, vol. 58, pp. 73–93, 2015.

[23]. Radhika Rani Chintala, Narasinga Rao M R, Somu Venkateswarlu, "Performance Metrics and Energy Evaluation of a Lightweight Block Cipher in Human Sensor Networks," International Journal of Advanced Trends in Computer Science and Engineering, Volume 8, No.4, pp. 1487-1490, 2019.

[24]. Arjen K. Lenstra, "Key Lengths", The Handbook of Information Security, 06/2004.

[25]. Kaliski B., Moore's Law. In: van Tilborg H.C.A., Jajodia S. (eds) Encyclopedia of Cryptography and Security. Springer, Boston, MA. https:// doi.org/10.1007/978-1-4419-5906-5_420m, 2011.

[26]. Philippe Garrault and Brian Philofsky, "HDL Coding Practices to Accelerate Design Performance," Virtex-4, Spartan-3/3L, and Spartan-3E FPGAs, Xilinx WP231, Jan 2006.